\documentstyle[12pt]{article}
\begin{document}

\begin{flushright}
TIFR/TH/93-10
\end{flushright}
\bigskip
\bigskip
\begin{center}
\large {\bf Radiative Electroweak Parameters} \\
\bigskip
\bigskip
Probir Roy \\
Tata Institute Of Fundamental Research \\
\bigskip
\bigskip
{\em Talk given at the X DAE High Energy Physics Symposium held at the Tata
Institute, Bombay, Dec. 26-31, 1992.} \\
\end{center}
\bigskip
\bigskip

\noindent Abstract

The status of the present precision measurements of electroweak
observables is reviewed with specific reference to the radiative
parameters $S,T,U$ or equivalently $\epsilon_1,\epsilon_2,\epsilon_3$.
The significance of the obliqueness hypothesis is underlined and the
importance of the ``local fit'' method of extracting these parameters from
the data is emphasized.  Possible new physics implications are briefly
touched upon.

\newpage

\noindent{\bf 1. Radiative/Precision Parameters}

The precision-testing of a renormalizable relativistic quantum field
theory is intrinsically connected with the accurate calculation of
radiative corrections in it.  This connection has a long tradition [1] in
QED.  The basic point there is extremely simple at the 1-loop-level.  If
one is measuring a quantity of order unity to the precision of the third
decimal place, comparison with theory is meaningful only if there is a
correct calculation taking all $0(\alpha)$ terms into account.
For the electroweak processes, being studied on the $Z$ mass-shell at LEP
1, the relevant couplings are semiweak.  Since the accuracy in the LEP
measurements  has now reached the $10^{-3}$ level, the consideration [2] of
1-loop corrections to the tree-level contributions is very pertinent.

Of course, certain couplings, relevant to weak and electromagnetic
processes, are known to an extraordinarily high accuracy.  Specifically, we
refer to two which form part of our reference frame : (1) the atomic fine
structure constant $\alpha_{EM}$, as measured in the A.C. Josephson
effect, namely [3]
$$
\alpha^{-1}_{EM} = 137.0359895(61);
$$
(2) the Fermi constant $G_\mu$, as measured in muon decay.  The latter
enters the muon lifetime $\tau_\mu$ via the formula
$$
\begin{array}{c}
\tau^{-1}_\mu = (192\pi^3)^{-1} G^2_\mu m^5_\mu f(m_e/m_\mu) \left(1 +
{3\over5} m^2_\mu M^{-2}_W\right) \\
{}~~~~~~~~~~~~~~~~~~~~~~~~~~~~~~~~~~~~~~~~\cdot \left[1 + (2\pi)^{-1}
\alpha_{EM}
(m^2_\mu) \left({25 \over 4} - \pi^2\right)\right].\\
\end{array}
\eqno (1)
$$
In (1) $m_\ell$ is the mass of the lepton of type $\ell$, $M_W$ is the
mass of the $W$,
$$
f(x) \equiv 1 - 8x + 8x^3 - x^4 - 12x^2\ell n x,
\eqno (2)
$$
and -- accounting for the running $\alpha^{-1}_{EM}$ from $eV$ energies to
the electron mass and eventually to the muon mass --
$$
\alpha^{-1}_{EM} (m^2_\mu) = \alpha^{-1}_{EM} - {1 \over 3\pi} \ell n
{m^2_\mu \over m^2_e} + {1 \over 6\pi} \simeq 136.
\eqno (3)
$$
Using (1), (2) and (3) in conjunction with the experimental value of
$\tau_\mu$, one can deduce that [3]
$$
G_\mu = 1.166389(22) \times 10^{-5} ~{\rm GeV}^{-2}.
$$

Evidently, our knowledge of $\alpha^{-1}_{EM}$ and $G_\mu$ is pretty
precise.  The unification of weak and electromagnetic interactions in the
Standard Model (SM) implies a relation between these two quantities,
namely
$$
G_\mu = \pi \alpha_{EM} \left[\sqrt{2} M^2_W \left(1 - M^2_W
M^{-2}_Z\right) (1 - \Delta r)\right]^{-1}.
\eqno (4)
$$
In (4) $\Delta r$ is a purely radiative constant, i.e. it vanishes in the
tree approximation.  To one loop, there are no QCD effects and a complete
perturbative electroweak calculation of $(\Delta r)_{1-{\rm loop}}$ has
been done [4].  However, at present, that cannot be used for a precision
test of the SM via (4).  Thus is because the result depends on the two yet
unknown masses $m_{t,H}$, i.e. those of the top quark and of the Higgs
scalar.  The former enters through the diagram of Fig. 1a and the latter
(taking the single Higgs doublet Minimal Standard Model) through that of
Fig. 1b.  The precision testing of the MSM will be possible by use of (4)
once $m_{t,H}$ get known accurately from direct measurements.

\vskip 2in

Next we come to the $\rho$-parameter.  For low energy measurements it can be
obtained (in the approximation of neglecting momentum transfers and lepton
masses) in terms of the charged current cross section $\sigma (\nu_\mu e
\rightarrow \mu \nu_e)$ divided by the neutral current cross section
$\sigma(\nu_\mu e \rightarrow \nu_\mu e)$ at the same energy:
$$
\rho = \left[{\sigma(\nu_\mu e \rightarrow \nu_\mu e) \left({1\over4} -
s^2_\theta + {4\over3} s^4_\theta\right)^{-1} \over \sigma(\nu_\mu e
\rightarrow \mu \nu_e)}\right]^{1/2}.
\eqno (5)
$$
In (5) $s^2_\theta \equiv 1 - M^2_W/M^2_Z = 1 - c^2_\theta \simeq 0.23$.
The present best experimental number for this parameter is [5]
$$
\rho = 1.008 \pm 0.014.
$$
In SM, for any number of Higgs doublets, $\rho$ is unity at the tree level
but changes radiatively.  The theoretical calculation is scheme-dependent.
In the on-shell renormalization scheme [6], the 1-loop expression contains
a quadratic dependence on $m_t$.  For $m_t \gg M_Z$ one can write
$$
\rho \simeq 1 + {3\alpha_{EM} \over 16\pi} {m^2_t \over s^2_\theta
c^2_\theta M^2_Z}.
\eqno (6)
$$

There is, however, a somewhat different $\rho$-parameter which pertains to
the on-shell $Zf\bar f$ coupling as probed in LEP 1.  The tree-level
$Zf\bar f$ vertex of the SM, with $T_{3f}$ referring to the $SU(2)_L$
3rd-component of isospin of $f$, is
$$
V^{\rm tree}_\mu = \sqrt{\sqrt{2} G_\mu} M_Z \gamma_\mu \left[(T_{3f} -
s^2_\theta Q_f) - T_{3f} \gamma_5\right].
$$
After 1-loop radiative effects it changes to
$$
V^{\rm 1-loop}_\mu = \sqrt{\sqrt{2} G_\mu} M_Z \gamma_\mu
\left[\sqrt{\rho^{EFF}} (T_{3f} - \sin^2 \theta^{EFF}_W Q_f) -
\sqrt{\rho^{EFF}} T_{3f} \gamma_5\right].
\eqno (7)
$$
High statistics measurements on the $Z$ at LEP 1 imply [7]
$$
\begin{array}{lcl}
\rho^{EFF} & = & 1.000 \pm 0.0036, \\[2mm]
\sin^2 \theta^{EFF}_W & = & 0.2324 \ \ (11).
\end{array}
$$
Allowing for the evolution of the fine structure constant $\alpha_{EM}$
from $eV$ energies to LEP 1, where [7]
$$
\alpha^{-1}_{EM} (M^2_Z) = 128.2 \pm 0.09^{+0.0m_t}_{-0.4m_t},
$$
(4) changes to
$$
G_\mu = \pi \alpha_{EM} (M^2_Z) \left[\sqrt{2} M^2_W (1 - M^2_W/M^2_Z) (1
- \Delta r_W)\right]^{-1}.
\eqno (8)
$$
The parameter $\Delta r_W$, appearing in (8), is also fully known to
1-loop and has a quadratic dependence on $m_t$.

The radiative parameters $\epsilon_1,\epsilon_2,\epsilon_3$ or $S,T,U$ can
now be defined as
$$
\epsilon_1 \equiv \rho^{EFF} - 1 \equiv \alpha_{EM} T,
\eqno (9a)
$$
$$
\begin{array}{lcl}
\epsilon_2 \equiv   c^2_\theta (\rho^{EFF} - 1) + s^2_\theta \Delta r_W
(c^2_\theta - s^2_\theta)^{-1} &-& 2\left(\sin^2 \theta^{EFF}_W\right)^{-1}
\\ [2mm]
&\equiv& - (4s^2_\theta)^{-1} \alpha_{EM} U, \\
\end{array}
\eqno (9b)
$$
$$
\begin{array}{lcl}
\epsilon_3  \equiv  (c^2_\theta - s^2_\theta) \left\{s^2_\theta \sin^2
\theta^{EFF}_W)^{-1} - 1\right\} &+& c^2_\theta (\rho^{EFF} - 1) \\[2mm]
&\equiv& (4s^2_\theta)^{-1} \alpha_{EM} S. \\
\end{array}
\eqno (9c)
$$
In motivating these strange-looking combinations, one may say the
following.  The $\epsilon_1$ parameter is just the radiative/new physics
part of the effective $\rho$-parameter on the $Z$ mass-shell, while the
combinations standing for $\epsilon_{2,3}$ have been chosen [8] in such a
way that the quadratic terms in $m_t$ cancel and only an insensitive
logarithmic dependence on the top mass survives.  Moreover, as explained
below, these combinations are the most natural 1-loop radiative parameter
in the ``obliqueness'' approximation [9] of retaining only vector boson
polarization terms and neglecting vertex corrections and box graphs.

\vskip 2in
\bigskip

\noindent {\bf 2. Obliqueness and oblique parameters}

\nobreak
In order to understand the efficacy of the obliqueness approximation, it
is instructive to look at the 1-loop terms in the muon decay amplitude.
The tree diagram involves the exchange of a charge-carrying $W$ between
the muon and the electron converting them into $\nu_\mu$ and $\bar \nu_e$
respectively.  At the 1-loop level, separately there are the $W$ vacuum
polarization contributions (Fig. 2a), vertex corrections for the $\mu
\rightarrow \nu_\mu$ transition (Fig. 2b), vertex corrections for the
$\nu_e \rightarrow e$ transition (Fig. 2c) and box-type graphs (Fig. 2d).
The vector boson propagator, to one loop, has in fact been enumerated in
Fig. 3, though of course the last tadpole graph is absent in the unitary
gauge and -- in any event -- drops out of ~renormalized ~on-shell
{}~amplitudes.

\newpage

\vspace*{2in}

\noindent These vacuum polarization contributions form a
gauge-invariant subset and will henceforth be called oblique corrections.
They totally dominate over the vertex corrections and the box graphs (by
nearly an order of magnitude) in their contributions to $\Delta r$.  This
numerical domination by the vacuum polarization terms is a generic feature
of all 1-loop physically interesting radiative corrections considered to
date with one important exception.  The latter is the $Zb\bar b$ coupling
where the vertex correction from a triangular loop with the two top and
one longitudinally polarized $W_L$ internal lines (Fig. 4) makes a
numerically significant contribution on account of the top-antitop-Higgs
coupling which enters in the $t\bar t W_L$ vertex.

\vskip 1.5in

Treating the $Zb\bar b$ vertex separately, one is then justified at least
at the 10\% level in keeping only the oblique corrections and ignore the
rest.  This causes a tremendous simplification in the problem as detailed
below.  All radiative effects to 1-loop can now be described in terms of
vacuum polarization terms that are gauge-independent $\Pi$-functions, i.e.
$$
\int d^4 x e^{iq.x} \langle\Omega|J^A_\mu(x) J_\nu(0)|\Omega\rangle =
-\Pi^{AB} (q^2) \eta_{\mu\nu} + q_\mu q_\nu ~{\rm terms}.
$$
The parameters $S,T,U$ are, in fact, linear combinations of
appropriately defined $\Pi$-functions.  As a result, they represent
compact, model-independent parametrizations of 1-loop radiative
corrections in the obliqueness approximation.  They are not only
gauge-independent but are renormalization scheme invariant since they
appear [10] in the coefficients of higher dimensional operators of a
1-loop effective Lagrangian density.  We shall see that there are two
additional desirable features of these parameters.  First, new physics
contributions to them add linearly to those from the SM; this actually is
a property of the $\Pi$-functions as can be seen by inserting a complete
set of states.  Second, they are optimal probes of any nondecoupled heavy
new physics, if present.  These points will be elaborated below.

The oblique parameters $S,T,U$ of (9) theoretically emerge from the
(generally divergent) $\gamma,Z$ and $W$ self-energies and the $\gamma -
Z$ mixing amplitudes $\Pi_{\gamma\gamma} (q^2)$, $\Pi_{ZZ} (q^2)$ and
$\Pi_{\gamma Z} (q^2)$ respectively (Fig. 5).  The latter are defined

\vskip 1.5in

\noindent as functions of the four momentum scale $q$ of the relevant gauge
bosons.
Electromagnetic gauge invariance implies $\Pi_{\gamma\gamma} (0) = 0 =
\Pi_{\gamma Z} (0)$.  Denote the weak isospin currents as $J^\mu_{1,2,3}$
and the electromagnetic current as $J^\mu_Q = J^\mu_3 + {1\over2}
J^\mu_Y$.  Thus the $Z$-current is $(e/s_\theta c_\theta) (J^\mu_3 -
s^2_\theta J^\mu_Q)$ where $e^2 = 4\pi \alpha_{EM}$.  Thus
$$
\Pi_{\gamma\gamma}  =  e^2 \Pi_{QQ}, \eqno (10a)
$$
$$
\Pi_{ZZ} =  e^2 s^{-2}_\theta c^{-2}_\theta (\Pi_{33} - 2s^2_\theta
\Pi_{3Q} + s^4_\theta \Pi_{QQ}), \eqno (10b)
$$
$$
\Pi_{WW}  =  e^2  s^{-2}_\theta \Pi_{11}, \eqno (10c)
$$
$$
\Pi_{\gamma Z} =  e^2 c^{-1}_\theta s^{-1}_\theta (\Pi_{3Q} - s^2_\theta
\Pi_{QQ}) \eqno (10d)
$$
at all values of $q^2$.  The ``theoretical'' definitions of $S,T$ and $U$
are
$$
\begin{array}{lcl}
S & = & 16\pi M^{-2}_Z \left[\Pi_{33} (M^2_Z) - \Pi_{33} (0) -
\Pi_{3Q} (M^2_Z)\right] \\[2mm]& = & 8\pi  M^{-2}_Z \left[\Pi_{3Y} (0)
- \Pi_{3Y} (M^2_Z)\right], \hfill (11a) \\[2mm]
T & = & 4\pi s^{-2}_\theta c^{-2}_\theta M^{-2}_Z \left[\Pi_{11} (0) -
\Pi_{33} (0)\right], \hfill (11b) \\[2mm]
U & = & 16\pi M^{-2}_W \left[\Pi_{11} (M^2_W) - \Pi_{11} (0)\right] -
16\pi M^{-2}_Z \left[\Pi_{33} (M^2_Z) - \Pi_{33} (0)\right]. \hfill
(11c)
\end{array}
$$
(11) and (9) match in the obliqueness approximation.

$T$ and $U$ receive nonzero contributions from the violation of weak
isospin and are finite on account of the weak isospin symmetric nature of
the divergence terms.  $S$ originates from the mixing between the weak
hypercharge and the third component of weak isospin as a consequence of
the spontaneous symmetry breakdown mechanism.  Soft operators, involved in
the latter, do not affect the leading divergence because of Symanzik's
theorem.  The nonleading divergence cancels out in the difference between
$\Pi_{3Y} (M^2_Z)$ and $\Pi_{3Y} (0)$, leaving a finite $S$.  The
$\Pi_{AB}$ functions receive contributions from different sources
additively.  This enables us to define $\tilde \Pi_{AB} = \Pi_{AB} -
\Pi^{SM}_{AB}$ where $\Pi^{SM}_{AB}$ is the standard model contribution.
Thus the twiddled pi function would arise purely from new physics beyond
SM.  The latter depends on the yet unknown Higgs mass $m_H$
(logarithmically, on the strength of Veltman's screening theorem) and the
top mass $m_T$ (with leading quadratic dependence).  In terms of direct
experimental searches as well as theoretical consistency in the
perturbative calculation scheme, one can say that $91 ~{\rm GeV} < m_T <
200~{\rm GeV}$ and $60~{\rm GeV} < m_H < 1~{\rm TeV}$.

Consistent with the practice in the recent literature [11,12], we choose
the SM reference point at $m_T = 140~{\rm GeV}$, $m_H = 100~{\rm GeV}$ and
the QCD fine structure constant on the $Z$ $\alpha_S (M_Z) = 0.120$.  Of
course, shifts in
these values can easily be incorporated [11].  The most reliable
extraction of $\tilde S,\tilde T,\tilde U$ is now from current accelerator
data.  The parameters $\tilde S$ and $\tilde T$ are best obtained by
performing a ``local fit'' [11] to various cross sections and asymmetries
at LEP for processes $e^+e^- \longrightarrow f\bar f$ as functions of the
CM energy $\sqrt{s}$ in the $Z$ lineshape region.  The original local fit
was done [11] with $700,000$ data points, but on updated analysis [13] with 1.5
million events around the $Z$ peak yields $\tilde S = -0.48 \pm 0.45$,
$\tilde T = -0.19 \pm 0.41$.  If one combines these with the rather
inaccurately
known value of $M_W$, then the use of (8) and (9b) leads to $U = -0.12 \pm
0.90$.  The errors will be significantly reduced once the $W$-mass is
better known.  In Fig. 5 we show the 90\% confidence level allowed
elliptic region in the $\tilde S,\tilde T$ plane.  It
should be pointed
out that most technicolor and walking technicolor scenarios (and in
general condenstate models of electroweak symmetry breaking), pertaining
to nondecoupled new physics,  predict large
positive values of $\tilde S$ and $\tilde T$ outside this ellipse and
are disfavored [14] by the data.  In constrast, supersymmetric models,
which stand for decoupled new physics,  generally
predict numerically small values of $\tilde S$ and $\tilde T$ close to
zero which cannot be tested at the present level of accuracy.

In conclusion, the radiative electroweak parameters $\epsilon_{1,2,3}$ (or
$S,T,U$) constitute compact, model-independent probes into new physics.
The contributions from the latter (marked by twiddles) are linearly
additive to those from the SM in the obliqueness approximation.  The data
tend to lie in the third quadrant
of the $\tilde S,\tilde T$ plane, disfavoring technicolor and related
condensate models.  The errors in $\tilde U$ will remain large until the
$W$-mass is determined to much better accuracy.

\vskip 6in
\bigskip

\noindent {\bf Acknowledgements}

I thank Sunanda Banerjee, Gautam Bhattacharyya, Francis Halzen, Ernest Ma,
Bill Marciano and Xerxes Tata for many stimulating discussions of these
issues.

\newpage

\begin{center}
{\bf References} \\
\end{center}

\begin{enumerate}
\item{} T. Kinoshita, ``Recent Developments of QED'', {\it Proc. 19th
Int. Conf. on High Energy Physics}, Tokyo (1978, eds. S. Homma {\it et
al}).

\item{} M. Veltman, Nucl. Phys. {\bf B123} (1977) 89.  A. Sirlin, Phys.
Rev. {\bf D22} (1980) 971.  W.J. Marciano and A. Sirlin, Phys. Rev. {\bf
D22} (1980) 2695; Errtm. {\it ibid}. {\bf D31} (1985) 213.

\item{} M. Aguilar-Benitez {\it et al}, Particle Data Group, Phys. Rev.
{\bf D45}, Part 2 (June 1992).

\item{} Z. Hioki, Mod. Phys. Lett. {\bf A7} (1992) 1009.

\item{} L. Rolandi, talk given at the {\it 26th Int. Conf. on High
Energy Physics} (Dallas, 1992).

\item{} W.F.L. Hollik, Fortschr. Phys. {\bf 38} (1990) 165.  F.
Jegerlehner, ``Physics of Precision Experiments with Z's'' in {\it
Progress in Particle and Nuclear Physics} (ed. A. F\"assler, Pergamon).

\item{} A. Gurtu, these proceedings.

\item{} G. Altarelli and R. Barbieri, Phys. Lett. {\bf B253} (1991) 161.
G. Altarelli, R. Barbieri and S. Jadach, Nucl. Phys. {\bf B369} (1992) 3.

\item{} D.C. Kennedy and B.W. Lynn, Nucl. Phys. {\bf B322} (1989) 1.
B.W. Lynn, SLAC-PUB-5077 (1989, unpublished).  D.C. Kennedy, Nucl. Phys.
{\bf B351} (1991) 81.

\item{} B. Holdom, Phys. Lett. {\bf B259} (1991) 329.  H. Georgi, Nucl.
Phys. {\bf B363} (1991) 301.

\item{} G. Bhattacharya, S. Banerjee and P. Roy, Phys. Rev. {\bf D45}
(1992) R729; Errtm. {\it ibid}. {\bf D46} (1992) 3215.

\item{} G. Altarelli, talk given at Recontres de Moriond on Electroweak
Interactions and Unified Theories (1992), CERN Report TH-6525-92, to be
published.

\item{} M.E. Peskin, talk given at the 26th Conf. {\it High Energy
Physics} (Dallas, August 1992).

\item{} J. Ellis, G.L. Fogli and E. Lisi, CERN Report TH-6643-92, to be
published; Phys. Lett. {\bf B285} (1992) 238.

\item{} R. Barbieri, M. Frigeni, F. Giuliani and H.E. Haber, Nucl.
Phys. {\bf B341} (1990) 309.  R. Barbieri, M. Frigeni and C. Caravaglios,
Phys. Lett. {\bf B279} (1992) 169.
\end{enumerate}

\end{document}